\documentclass[aip, amsmath, amssymb, reprint, 9pt]{revtex4-2}
\usepackage{bm}
\usepackage[title]{appendix}
\usepackage{graphicx, tikz, orcidlink}
\usepackage{epstopdf}
\usepackage{ascmac}
\usepackage{ulem}
\usepackage{cancel}
\usepackage{here}
\usepackage{physics}
\usepackage{bm}
\usepackage{comment}

\begin{document}


\title{Quantum Hierarchical Fokker--Planck Equations with U(1) Gauge Fields (U(1)-QHFPE): A Computational Framework for Aharonov--Bohm Effects
}
\date{Last updated: \today}
\author{Shoki KOYANAGI\orcidlink{0000-0002-8607-1699}}\email[Author to whom correspondence should be addressed: ]{koyanagi.shoki.e00@kyoto-u.jp}
\author{Hyeonseok YANG\orcidlink{0009-0007-1308-4724}}
\author{Yoshitaka TANIMURA\orcidlink{0000-0002-7913-054X}}
\affiliation{Department of Chemistry, Graduate School of Science,
Kyoto University, Kyoto 606-8502, Japan}

\begin{abstract}
We introduce U(1)-QHFPE, a non-Markovian and non-perturbative open quantum dynamics software package for solving quantum Fokker--Planck equations incorporating gauge fields within the Hierarchical Equations of Motion (HEOM) formalism. The framework rigorously preserves gauge invariance and rotational symmetry, enabling accurate simulations of transport phenomena such as the Aharonov--Bohm effect under strong system-bath coupling. In this regime, quantum entanglement between the system and bath emerges naturally. Demonstration programs include calculations of response functions in Aharonov--Bohm ring geometries, showcasing the code's ability to resolve topological quantum interference in dissipative open systems.
\end{abstract}

\pacs{}

\maketitle

\section{Introduction}\label{sec:intro}

Gauge invariance plays a foundational role in modern theoretical physics, serving as a guiding principle that determines the structure of physical laws.\cite{Weinberg1995,Coleman2019} 
By requiring that observable quantities remain invariant under local transformations of the underlying fields,  gauge invariance ensures that only observables, such as electric and magnetic fields, appear in the theory, filtering out unphysical degrees of freedom.
According to Noether’s theorem, gauge symmetries are directly linked to conservation laws; for instance, U(1) symmetry implies charge conservation and naturally gives rise to interaction fields, with the photon emerging as the gauge boson of the U(1) group. Moreover, gauge invariance underlies the coherence of quantum phenomena, as exemplified by the Aharonov--Bohm (AB) effect,\cite{1959Aaronov-Bohm,Tonomura1986PhysRevLett.56.792,2008NoriTonomura} where nonlocal phase shifts manifest in measurable interference patterns.

While the fundamental aspects of gauge-induced quantum effects have been extensively studied across various platforms, the influence of dissipation and thermal fluctuations, particularly those mediated by phonons, on quantum coherence remains incompletely understood.\cite{2015GreenDissipationLi} In metallic and semiconductor nanorings, gate-induced resistance is often dominant, whereas in molecular nanorings, intrinsic resistance within the conductor becomes more relevant.\cite{2018GreenPhysRevB.98.125417,2020ODissipationdoi:10.1021/acs.jpcc.0c01706,2011ABdephPhysRevB.84.035323,2012ABdepha2dotsPhysRevB.86.115453,2016ABdephaLIU2016163} Although AB rings have been investigated through diverse configurations, including multi-ring networks,\cite{1986ABPhysRevLett.56.386} embedded quantum dots,\cite{1995AB_dotPhysRevLett.74.4047,1995AB_dotPhysRevB.52.R14360,2002AB_dotPhysRevLett.88.256806,KATSUMOTO2003151,Ando2004PhysRevB.69.115307,Tanatar2005PhysRevB.71.125338,Li2025} multiterminal geometries,\cite{2009ABMultiterminalPhysRevB.79.195443} and tunneling junctions,\cite{1993ABtunnelPhysRevB.47.2768} the role of dissipation in these systems remains elusive.

Phonon-based models have been employed to analyze AB rings,\cite{ABYiJinJIn_2018,2019CaoPhysRevB.99.075436} but they offer limited insight into gauge invariance, as they do not inherently encode the topological structure of electromagnetic potentials. Geometric features can be described using coordinate-based models such as the Caldeira–Leggett (CL) framework,\cite{Caldeira1981,Caldeira1983,Leggett1984PhysRevB.30.1208,Chen1986,Leggett2001} yet when applied to systems with rotational symmetry, such as quantum rotors and AB rings, these models fail to preserve entanglement under rotation due to the symmetry-breaking nature of the thermal bath, yielding only semiclassical behavior.\cite{ST02JPSJ,ST03JCP}

In our previous work,\cite{YKT25JCP1} we demonstrated that the introduction of a rotationally invariant system–bath (RISB) model is crucial for preserving quantum entanglement in rotating systems coupled to a thermal environment.\cite{IT18JCP,IT19JCP,Lipeng3Drotar2019} Building on this foundation, we developed a gauge-invariant formalism based on the HEOM, referred to as U(1)-HEOM and its Fokker-–Planck variant, U(1)-QHFPE, which is applicable to strongly coupled, non-Markovian open quantum systems. While the three-dimensional (3D) U(1)-HEOM formulation is computationally intensive, its reduction to two-dimensional (2D) geometries, such as AB rings, via U(1)-QHFPE yields substantial simplification. The present work extends our previous implementation, enhancing its generality and releasing the software as a publicly accessible tool for the broader research community.

Recent interest in quantum systems sensitive to gauge fields has grown rapidly,\cite{FrancoNORI2019resolution,FrancoNORI2020gauge, FrancoNORI2021gauge,FrancoNORI2021QEDPhysRevResearch.3.023079} particularly in cavity-quantum electrodynamics (cavity-QED) platforms, where quantum optical master equations are often employed under Markovian and factorized assumptions.\cite{FrancoNORI2022gauge,FrancoNORI2023generalized} In contrast, the U(1)-QHFPE framework developed here enables the study of such systems under realistic conditions, capturing thermal excitation and dissipation effects at ultralow temperatures where quantum coherence is essential. As part of this work, we extend previous results by incorporating symmetric and antisymmetric correlation functions in non-Markovian and non-perturbative domains, and further explore AB phase oscillations.

The remainder of this paper is organized as follows. Section \ref{sec:theory} presents the theoretical models and derivation of the U(1)-QHFPE. Section 
\ref{sec:soft} explains the detail of software.  Sec. \ref{sec:ND} demonstrates the calculated results of response functions for the AB ring. Section \ref{sec:Conclution} provides concluding remarks.

\section{Hierarchical Equations of Motion for AB ring system}
\label{sec:theory}
We consider the AB system described as the 2D RISB model expressed as $
\hat{H}_{\rm tot} ( t ) = \hat{H}_A ( t ) + \hat{H}_{I + B} $, where $\hat{H}_A ( t )$ is the Hamiltonian of the AB system defined as
\begin{align} 
\label{eq:HA}
\hat{H}_A ( t ) = \frac{1}{2 I_S} ( \hat{p}_\theta - e r_0 A_\theta )^2 + U ( \hat{\theta} ; t ) ,
\end{align}
Here $\hat{p}_\theta , e, r_0$ and $A_\theta$ are the 2D angular momentum operator of the particle, electric charge, radius of the ring and the azimuthal component of the vector potential, and $I_S = m_S r_0^2$ is the moment of inertia of the particle.  We assume that the vector potential is time-independent and the constant in the $\theta$ direction. We also introduce the potential, $U ( \hat{\theta} ; t )$, where $\hat{\theta}$ is the angular position operator. The system--bath (S-B) interaction plus bath Hamiltonian $\hat{H}_{\mathrm{I} + \mathrm{B}}$ is expressed as
\begin{eqnarray}
\hat{H}_{I + B} &&\nonumber \\
= \sum_{\alpha}^{x,y}&& \sum_k \left\{
\frac{(\hat{p}_{k}^{\alpha})^2}{2 m_k^{\alpha}}  + \frac{m_k^{\alpha}  (\omega_k^{\alpha})^2}{2} \left(\hat{q}_k^{\alpha}  -
\frac{c_{k}^{\alpha} \hat V_{\alpha}}{m_{k}^{\alpha} (\omega_{k}^{\alpha})^{2}} \right)^2\right\}, \nonumber \\
\label{eq:HIB}
\end{eqnarray}
where $\hat{p}_j^\alpha , \hat{x}_j^\alpha , m_j^\alpha$ and $\omega_j^\alpha$ are the momentum and position operators , mass, and angular frequency of the $j$th mode of the bath in the $\alpha$ direction. The coefficient $c_j^\alpha$ is the coupling constant between the system and the $j$th mode of the $\alpha$-direction bath. We set the system operator, $\hat{V}_\alpha$, as $\hat{V}_x = r_0 \cos \hat{\theta}$ and $\hat{V}_y = r_0 \sin \hat{\theta}$. We include the counter terms that are introduced to maintain the translational symmetry of the system.\cite{TW91PRA} The harmonic bath  in the $\alpha$ direction  is characterized by the spectral distribution function,  defined as
$J^{\alpha}(\omega) =\sum_k [{{\hbar}( c_k^{\alpha})^2}/{2m_k^{\alpha} \omega_k^{\alpha}}] \delta(\omega - \omega_k^{\alpha})$,
and the inverse temperature, $\beta \equiv 1/k_{\mathrm{B}}T$, where $k_\mathrm{B}$ is the Boltzmann constant. 

To easily adapt the HEOM formalism, we use the Drude spectral density expressed as\cite{T06JPSJ,T20JCP}
\begin{align}
\label{eq:DrudeSDF}
J_\alpha ( \omega ) = \frac{\hbar \eta_\alpha}{\pi} 
\frac{\gamma_\alpha^2 \omega}{\gamma_\alpha^2 + \omega^2} ,
\end{align}
where $\gamma_\alpha$ and $\eta_\alpha$ are the inverse of the noise correlation time and S-B coupling strength in the $\alpha=x$ and $y$ direction. It should be noted that $ J^{\alpha} (\omega)$ does not have to be identical for different $\alpha$.

We  introduce the discretized Wigner transformation with periodic boundary conditions (DWT-PBC).\cite{YKT25JCP1}  For the density operator in the 2D periodic system, this is defined as
\begin{align}
\label{eq:SystemWDF}
& W_{\{ \bm{n}_\alpha \}} ( p_n , x ; t ) \nonumber \\
&
~~~~~~~~~= \frac{1}{2 \pi \hbar} \int^{L}_{- L} e^{\frac{i p_n \varphi}{\hbar}}
\rho_{\{ \bm{n}_\alpha \}}  \left( x - \frac{p_n}{2} ,x + \frac{p_n}{2} ; t \right) d \varphi,
\end{align}
where $\rho_{\{{\bm n}_{\alpha}\}}\left(x, {x}' ; t\right)$ and $W_{\{ \bm{n}_\alpha \}} ( p_n , x ; t )$ are an auxiliary density operator (ADO) and an auxiliary Wigner distribution function (WDF) in the HEOM formalism, and $L$ is the period of the system, set to $2 \pi$ for the AB ring system.  The subscript vectors are, for example, set as $\{ \bm{n}_\alpha \} = \{ \vec{n}_x , \vec{n}_y \}$ with $ \bm{n}_\alpha \equiv ( n_\alpha^0 , n_\alpha^1 , \cdots , n_\alpha^{K_\alpha} )$, representing a set of non-negative integers.
The function $W_{\{ \bm{n}_\alpha \}} ( p_n , x )$ is obtained for the periodic boundary condition. The momentum variable is discretized as $p_n = n \hbar / 2$, where $n$ is an integer. The WDF associated with the zero-index vectors, i.e., $\bm{n}_x = \bm{0}$ and $\bm{n}_y = \bm{0}$. 

In the DWT-PBC representation, the U(1)-hierarchical quantum Fokker--Planck equations (U(1)-HQFPE) are expressed as\cite{YKT25JCP1}
\begin{align}
\label{eq:HQFPE}
&
\frac{\partial}{\partial t} W_{\{ \bm{n}_\alpha \}} ( p_n , \theta ; t ) \nonumber \\
&
~~~~~~= - \left( \hat{\mathcal{L}}_{qm} + 
\sum_\alpha^{x , y} \sum_{j = 0}^{K_\alpha} n_\alpha^j \nu_j^\alpha \right) 
W_{\{ \bm{n}_\alpha \}} ( p_n , \theta ; t ) \nonumber \\
&
~~~~~~+ \sum_\alpha^{x , y} \sum_{j = 0}^{K_\alpha} \hat{\Phi}_\alpha 
W_{\{ \bm{n}_\alpha + \bm{e}_\alpha^j \}} ( p_n , \theta ; t ) \nonumber \\
&
~~~~~~+ \sum_\alpha^{x , y} \sum_{j = 0}^{K_\alpha} n_\alpha^j \hat{\Theta}_j^\alpha
W_{\{ \bm{n}_\alpha - \bm{e}_\alpha^j \}} ( p_n , \theta ; t ) ,
\end{align}
where $\nu_j^\alpha$ and $K_\alpha$ are the $j$th Pad{\'e} frequency and the number of the Pad{\'e} frequencies in the $\alpha$ direction. 
The operator $\hat{\mathcal{L}}_{qm}$ is the quantum Liouvillian expressed as
\begin{align}
\label{eq:defQuantumLiouvillian}
- \hat{\mathcal{L}}_{qm} W (p_n,\theta  )
&= - \frac{p_n - q r_0 A_\theta}{I_S} \frac{\partial W ( p_n,\theta  )}{\partial \theta} \nonumber \\
&
+ \frac{r_0^2 ( \eta_y \gamma_y - \eta_x \gamma_x )}{4 \hbar} \sin ( 2 \theta ) \nonumber \\
&
~~~~~ \times
\left( W (p_{n + 2},\theta ) - W ( p_{n - 2},\theta ) \right) .
\end{align}
Here, the second term in the right-hand side in Eq.~\eqref{eq:defQuantumLiouvillian} is the contribution from the counter term and vanishes in the isotropic case. The other operators are defined as $\hat{\Phi}_\alpha = r_0 f_\alpha ( \theta ) \delta / \delta p_n \quad ( \alpha = x , y )$,
\begin{align}
\label{eq:defTheta0}
\hat{\Theta}_0^\alpha W ( p_n , \theta )
&
= \frac{\eta_\alpha r_0 \gamma_\alpha}{\beta} 
\left( 1 + \sum_{k = 1}^{K_\alpha} \frac{2 \bar{\eta}_j^\alpha \gamma_\alpha^2}
{\gamma_\alpha^2 - ( \nu_j^\alpha )^2} \right) 
\frac{\delta W ( p_n , \theta )}{\delta p_n}
\nonumber \\
&
- \frac{\eta_\alpha r_0 \gamma_\alpha^2}{2} g_\alpha ( \theta )
\left( W ( p_{n + 1} , \theta ) + W ( p_{n - 1} , \theta ) \right) ,
\end{align}
and
\begin{align}
\label{eq:defThetaJ}
\hat{\Theta}_j^\alpha = - f_\alpha ( \theta )
\frac{\eta_\alpha r_0 \gamma_\alpha^2}{\beta} 
\frac{2 \bar{\eta}_j^\alpha \nu_j^\alpha}{\gamma_\alpha^2 - ( \nu_j^\alpha )^2}
\frac{\delta}{\delta p_n} \quad ( j = 1 , \cdots , K_\alpha ) ,
\end{align}
where $\bar{\eta}_j^\alpha$ is the $j$th Pad{\'e} coefficient in the $\alpha$ direction. The functions $f_\alpha ( \theta )$ and $g_\alpha ( \theta )$ are defined as $f_x ( \theta ) = - \sin \theta , f_y ( \theta ) = g_x ( \theta ) = \cos \theta$, and $g_y ( \theta ) = \sin \theta$, respectively. We also introduce the operator $\delta / \delta p_n$ defined as $\delta f ( p_n ) / \delta p_n \equiv ( f ( p_{n + 1} ) - f ( p_{n - 1} ) ) / \hbar$, where $f ( p_n )$ is an arbitrary function of $p_n$.

Although the U(1)-HQFPE contains the infinite number of the auxiliary WDFs, we introduce the truncation of the hierarchy depth, $N_{\rm max}$, and only consider the auxiliary WDFs with $N \leq N_{\rm max}$, where $N = \sum_\alpha^{x , y} \sum_{j = 0}^{K_\alpha} n_\alpha^j$.

\section{Software detials}
\label{sec:soft}
\subsection{Parallel Processing}
The provided software is compatible with both Central Processing Unit (CPU) and Graphics Processing Unit (GPU) architectures. As the U(1)-HQFPE in Eqs.~\eqref{eq:HQFPE}-\eqref{eq:defThetaJ} forms a hierarchy of simultaneous differential equations, each associated ADO can be propagated independently and in parallel. This hierarchical structure facilitates efficient parallelization of the computational workflow. CPU-based execution is realized via Open MultiProcessing (OMP) or Message Passing Interface (MPI),\cite{Schulten2012,Shi2009HEOM_MPI,Noack2018DMHEOM} while GPU acceleration leverages the Compute Unified Device Architecture (CUDA).\cite{Kramer2011,TT15JCTC,QuTiP_HEOM_GPU} Note that CUDA execution requires an NVIDIA GPU.

\subsection{Adaptive time step size method}

In our implementation, we adopt the Runge-Kutta-Fehlberg method~\cite{fehlberg1969low}, which dynamically adjusts the time step size $\Delta t$ to ensure that the numerical error remains below a user-defined tolerance parameter, TOL. Consequently, users are not required to manually specify an appropriate time step size. To estimate the local truncation error, we perform a single-step time evolution from $t$ to $t + \Delta t$ using both fourth- and fifth-order integration formulas, denoted as $W_{\{ \bm{n} \}}^{4\mathrm{th}}(p_n, \theta; t + \Delta t)$ and $W_{\{ \bm{n} \}}^{5\mathrm{th}}(p_n, \theta; t + \Delta t)$, respectively.

Because evaluating the error across all auxiliary WDFs is computationally expensive, we estimate the error using only the system WDF at the point $\theta = 0$. The estimated error is defined as
\begin{align}
\epsilon_{\rm err} &= \max_{p_n} \Big| W^{4th}_{\{ \bm{0} \}} ( p_n , \theta = 0 ; t + \Delta t ) 
\nonumber \\
&
~~~~~~~~~~~~- W^{5th}_{\{ \bm{0} \}} ( p_n , \theta = 0 ; t + \Delta t ) \Big| .
\label{eq:defEstimatedError}
\end{align}

If $\epsilon_{\mathrm{err}}$ exceeds TOL, the time step is updated according to
\begin{align}
\label{eq:NextTimeStepsize}
\Delta t_{\rm new} = \left( C \times {\rm TOL} / \epsilon_{\rm err} \right)^{0.2} \Delta t ,
\end{align}
where $C < 1$ is a safety factor introduced to control $\epsilon_{\mathrm{err}}$; in our implementation, we set $C = 0.99$. The time evolution is then retried using $\Delta t_{\mathrm{new}}$. If the estimated error is below TOL, the new time step is computed using Eq.~\eqref{eq:NextTimeStepsize}, and the simulation proceeds to the next time step.

\section{Numerical Demonstration}
\label{sec:ND}
\subsection{Numerical Details}

To demonstrate the numerical performance of our method, we consider the AB ring system with the following parameter settings: $m_S = 0.5$, $r_0 = 1.0$, $e = -1.0$, $\beta = 1.0$, and $\gamma_x = \gamma_y = 1.0$, which yields a moment of inertia $I_S = 1.0$. Here, we set $\hbar = 1$ and $k_{\rm B} = 1$. The external potential is set to zero, $U(\theta; t) = 0$.
The eigenenergies of the system are then expressed as $E_n = ( n - \bar{\Phi} )^2 \hbar \omega_0$, where  $n \in \mathbb{Z}$ and the characteristic frequency of the particle's rotational motion is defined by $\omega_0 \equiv \hbar / (2 I_S)$. Under the chosen parameters, we obtain $\omega_0 = 1.0.$
We introduce the dimensionless magnetic flux \( \bar{\Phi} \equiv \Phi / \Phi_0 \), where the enclosed magnetic flux within the AB ring is given by $\Phi \equiv A_\theta / (2 \pi r_0)$. The flux quantum is defined as \( \Phi_0 \equiv h / |e| \), representing the fundamental unit of magnetic flux quantization.
For numerical calculations, we set the tolerance parameter as ${\rm TOL} = 1.0 \times 10^{-10}$.

\subsection{Equilibrium Distribution}

\begin{figure}
\includegraphics[width=0.7\linewidth]{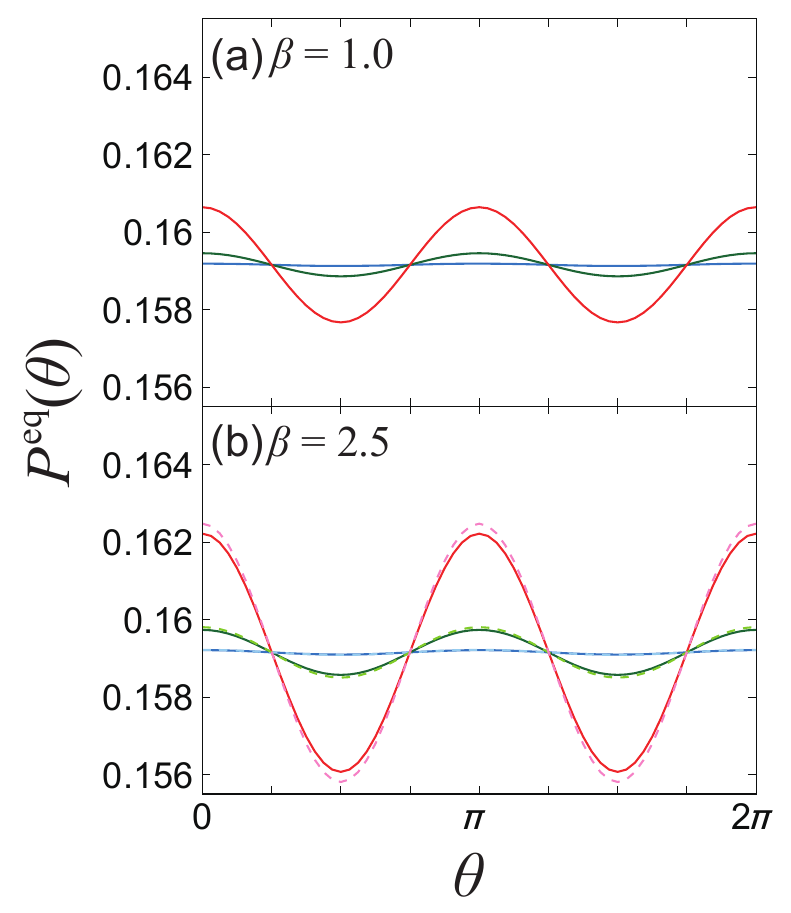}
\caption{\label{fig:Eq}Equilibrium PDFs are presented for (a) the high-temperature case with $\beta = 1.0$ and (b) the low-temperature case with $\beta = 2.5$. 
The S-B coupling strength in the $x$-direction was set to be twice that in the $y$-direction. At each temperature, three regimes of anisotropic S-B coupling are illustrated: weak coupling (blue curves, $\eta_x = 0.02$, $\eta_y = 0.01$), intermediate coupling (green curves, $\eta_x = 0.2$, $\eta_y = 0.1$), and strong coupling (red curves, $\eta_x = 1.0$, $\eta_y = 0.5$). In panel (b), the influence of a magnetic flux $\bar{\Phi} = 0.5$ is shown as dashed curves. At high temperature, however, the equilibrium distribution exhibits negligible dependence on the magnetic flux. Consequently, results for $\bar{\Phi} = 0.5$ are omitted from panel (a).}
\end{figure}

We begin by introducing the position distribution function (PDF), which is derived by integrating the system WDF over the momentum space. This procedure yields the spatial probability density along the angular coordinate $\theta$, and is formally defined as
\begin{align}
\label{eq:defPDF}
P ( \theta , t ) = \frac{\hbar}{2} \sum_{n = - \infty}^\infty W_{\{ \bm{0} \}} ( p_n , \theta ; t ) .
\end{align}

To illustrate the applicability of the developed code, we investigated a system coupled to an anisotropic bath. The S-B coupling strength in the $x$-direction was set to be twice that in the $y$-direction. We then considered three regimes: strong coupling ($\eta_x = 1.0$, $\eta_y = 0.5$), intermediate coupling ($\eta_x = 0.2$, $\eta_y = 0.1$), and weak coupling ($\eta_x = 0.02$, $\eta_y = 0.01$).

To perform the numerical integration of Eqs.~\eqref{eq:HQFPE}--\eqref{eq:defThetaJ}, the number of Pad\'e frequencies was set to $K_x = K_y = 2$ for the high-temperature case ($\beta = 1.0$) and $K_x = K_y = 4$ for the low-temperature case ($\beta = 2.5$). The truncation numbers were chosen as $N_{\mathrm{max}} = 7$, $N_{\mathrm{max}} = 6$, and $N_{\mathrm{max}} = 4$ for the strong, intermediate, and weak coupling regimes, respectively. Time evolution was continued until system WDF reached a steady state, at which point the resulting WDF was designated as the equilibrium distribution $W_{\{ \bm{0} \}}^{\mathrm{eq}}(p_n, \theta)$.

In Fig.~\ref{fig:Eq}, the equilibrium PDF is shown for the strong (red curve), intermediate (green curve), and weak (blue curve) S-B coupling cases under isotropic conditions, for (a) the high-temperature case ($\beta = 1$) and (b) the low-temperature case ($\beta = 2.5$).

In Fig.~\ref{fig:Eq}(a), since the heat bath in the $x$-direction couples more strongly to the system than that in the $y$-direction, $P^{\rm eq}(\theta)$ exhibits maxima at $\theta = 0$ and $\pi$. This is a purely quantum mechanical effect. In the classical limit, the distribution becomes the uniform equilibrium distribution ${\rm exp} [-\beta \hat{H}_S]$, which is independent of $\theta$, and no anisotropy is observed. Because the bath possesses 2D rotational symmetry, even when the coupling strengths in the $x$- and $y$-directions differ, the entanglement between the rotational and reduced radial degrees of freedom is essential to the emergence of anisotropy. Due to the high temperature, the effect of discretized momentum is suppressed, and the position distribution function shows almost no dependence on the magnetic field.

Figure~\ref{fig:Eq}(b) illustrates the low-temperature regime. The dashed curves correspond to the case with magnetic flux $\bar{\Phi} = 0.5$. In comparison with the high-temperature scenario, the amplitude is enhanced due to reduced decoherence from the bath, which lessens its disturbance of quantum entanglement. As the magnetic flux increases from $0$ to $0.5$, the oscillation amplitude grows, reflecting stronger quantum interference effects. According to the Byers--Yang theorem,\cite{ByersYang1961} physical observables in an AB ring, such as the oscillation amplitude or the persistent current, exhibit periodicity in $\bar{\Phi}$ with a unit period. Consequently, the amplitude diminishes in the flux interval $\bar{\Phi} \in [0.5, 1]$.

The attenuation of anisotropic entanglement serves as strong evidence supporting the existence of quantum S-B entanglement.

\subsection{Linear Response Function}

\begin{figure}
\includegraphics[width=8cm]{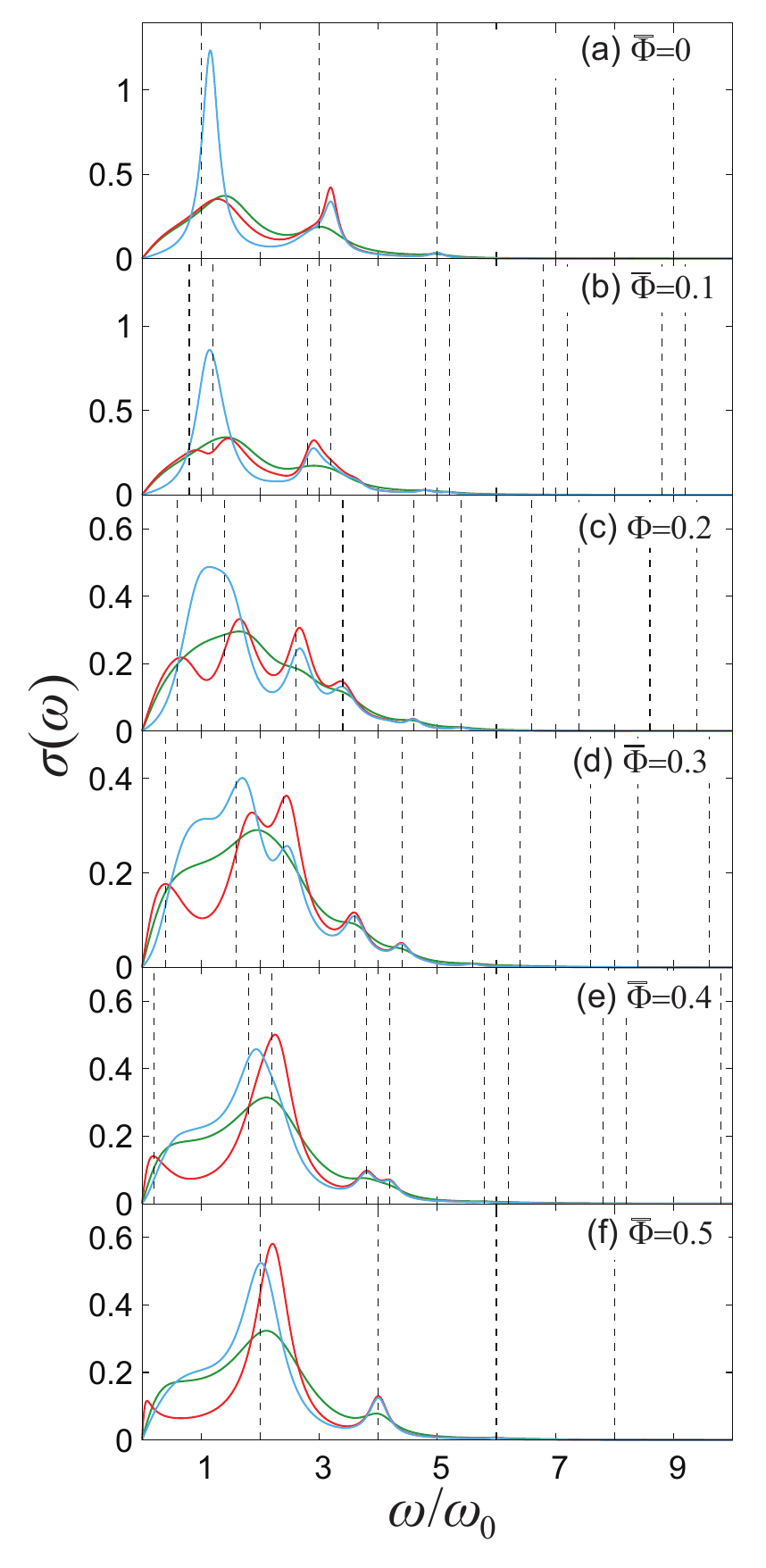}
\caption{\label{fig:LRF}Linear response spectra of the dipole moment for the AB ring, calculated under isotropic and anisotropic environments. 
The S-B coupling strengths are set as follows: 
    (red) $\eta_x = 1.0$, $\eta_y = 0.1$; 
    (green) $\eta_x = 1.0$, $\eta_y = 1.0$; 
    (blue) $\eta_x = 0.1$, $\eta_y = 1.0$.   The magnetic flux values for each panel are: 
    (a) $\bar{\Phi} = 0.0$, 
    (b) $\bar{\Phi} = 0.1$, 
    (c) $\bar{\Phi} = 0.2$, 
    (d) $\bar{\Phi} = 0.3$, 
    (e) $\bar{\Phi} = 0.4$, and 
    (f) $\bar{\Phi} = 0.5$. The black dashed line marks the peak position predicted by the system Hamiltonian without the baths.}
\end{figure}

In Fig.~\ref{fig:LRF}, we present the linear repose (rotational) spectrum defined as
\begin{align}
\sigma (\omega) = {\rm Im} \left\{\int_0^{\infty} dt \ {\rm e}^{i \omega t }
{R}^{(1)}(t) 
\right\}, \label{eqn:ProjCurr}
\end{align}
where the linear response function (LRF) of the dipole moment is defined as
${R}^{(1)}(t) = {i}\langle [\cos{\hat\theta(t )},  \cos{\hat\theta (0)} ] \rangle/{\hbar}$.
where
\begin{eqnarray}
\label{R_1t}
R^{( 1 )} ( t ) = \frac{i}{\hbar} {\rm Tr}_{\rm tot}
\left\{ \cos \hat{\theta} \hat{\mathcal{G}} ( t ) ( \cos \hat{\theta} )^\times \hat{\rho}_{\rm tot}^{\rm eq} 
\right\} 
\end{eqnarray}
and $\hat{\mathcal{G}}(t)$ is the time evolution operator for the total system in the absence of a laser interaction, and $\hat{\rho}_{\rm tot}^{\rm eq}$ is the equilibrium density operator for the total system. 

The LRF can be computed in the Wigner representation through the following procedure: 
(1) Apply the operator $\sin \theta \, (\delta / \delta p_n)$ to all auxiliary WDFs in the equilibrium state; 
(2) propagate the system under Eq.~\eqref{eq:HQFPE} from time $t = 0$ to $t$; 
(3) evaluate the LRF at time $t$ as the expectation value of $\cos \hat{\theta}$, given by
\begin{align}
\langle \cos \hat{\theta} \rangle \equiv \frac{\hbar}{2} \sum_{n = - \infty}^\infty \int^{2 \pi}_0 \cos \theta \;
W_{\{ \vec{0} \}} ( p_n , \theta ; t ) d \theta.
\end{align}

We set the SB coupling strength parameters as (red) $\eta_x = 1.0$ and $\eta_y = 0.1$, (green) $\eta_x = 1.0$ and $\eta_y = 1.0$, and (blue) $\eta_x = 0.1$ and $\eta_y = 1.0$. These parameter sets are applied to panels (a)–(f), each corresponding to a different magnetic flux $\bar{\Phi}$. 

Transitions induced by the external pulse are restricted to those between adjacent eigenstates of the system Hamiltonian, i.e., from the $n$th to the $n \pm 1$th states. The peak positions without baths in the spectrum can be estimated by evaluating the energy differences 
 $| E_n - E_{n \pm 1} |$  for $n \in \mathbb{Z}$. This leads to the analytical expression for the peak positions: $\omega / \omega_0 = 2 n + 1 \pm 2 \bar{\Phi}$. 
These predicted positions are indicated by black dashed lines in Fig.~\ref{fig:LRF}.

As $\bar{\Phi}$ increases, the peaks in the red curves merge and become and become progressively sharper [Fig.~\ref{fig:LRF}(d)-(f)]. This sharpening is a consequence of the enhanced ground-state particle velocity, resulting from the momentum shift induced by the vector potential. Accordingly, the particle shown in Fig.~\ref{fig:LRF}(e) and (f) couples less effectively to the bath along the $x$-direction, leading to a more pronounced spectral sharpening compared to the zero-field case.  Notably, the peak in the low-frequency region deviates from the dashed line due to the influence of S-B entanglement.\cite{ST02JPSJ,ST03JCP}

In contrast, under the blue curve configuration, the particle must migrate toward $\theta = 0$ and $\pi$ in order to interact effectively with the bath along the $y$-direction. The peaks in the blue curves broaden with increasing magnetic field strength. This broadening arises from the enhanced coupling between the particle and the $x$-direction bath, facilitated by the momentum shift induced by the vector potential.

At higher frequencies, however, the red and blue curves converge. This convergence arises because the spectral contributions in the high-frequency domain originate from high-momentum states, where the particle motion is sufficiently rapid. Under such conditions, the anisotropic nature of the bath coupling is effectively averaged out, resulting in similar spectral profiles irrespective of the directional coupling configuration.

\section{Conclusion}
\label{sec:Conclution} 
This paper enables numerically  ``exact'' dynamic simulations\cite{T20JCP} of AB rings under non-Markovian and non-perturbative thermal environments, from low temperatures, where quantum effects become significant, to high-temperature limits corresponding to the classical regime.  In addition, it accounts for anisotropic environmental effects, as demonstrated by the examples presented in this paper.

The Wigner description is optimal for device simulations because it not only incorporates environmental effects to satisfy fluctuation and dissipation conditions, but also facilitates geometric characterization and allows for the addition of periodic or inflow/outflow boundaries.\cite{Frensley1990}  

While this study focuses on isolated ring systems, verifying phase oscillations arising from the AB effect requires attaching leads to allow current flow. Such an extension is feasible through the generalization of the discrete Wigner transformation, which remains a subject for future investigation.

Furthermore, the proposed method is not limited to 1D systems; it is also applicable to electrons propagating in 2D geometries. This generality enables investigations into how dissipation and thermal effects influence phenomena such as the quantum Hall effect.

The Wigner representation facilitates the incorporation of time-dependent external fields of arbitrary strength and finds applications in quantum ratchet systems.\cite{KT13JPCB} It further accommodates arbitrary potential profiles, including those relevant to resonant tunneling systems.\cite{ST14NJP} Moreover, this framework can be extended to systems with multiple potential energy surfaces,\cite{IT17JCP} thereby allowing for the inclusion of gauge fields in Berry phase analyses.\cite{IT18CP}

The program is highly optimized and can run efficiently on personal computers. By promoting larger-scale parallelization and advanced GPU utilization, simulations such as those described above should become possible.

The fundamental principle of the U(1)-HEOM in gauge and rotational invariance lies in constructing the total Hamiltonian, including the thermal bath, in a manner that strictly preserves each symmetry. Therefore, equivalent results are expected from the pseudomode (PM) approach~\cite{Burghardt2011, Plenio2018, Lambert2019, Petruccione2020, pseudomodesNori2024}. This program may also serve as a reference when developing such programs. While simulating more realistic scenarios requires more computational resources than current classical hardware can provide, quantum computers may ultimately be required to perform such calculations~\cite{Dan2024qHEOM,Nori2024PRR,Shi2025MCEHEOM}.

Such extensions should be pursued as necessary, in step with ongoing advances in computational technology and algorithmic development.

\section*{Acknowledgments}
S. K.  was supported by Grant-in-Aid for JSPS Fellows (Grant No.24KJ1373). Y. T. was supported by JST (Grant No. CREST 1002405000170). H. Y. acknowledges a fellowship supported by JST SPRING, the establishment of university fellowships toward the creation of science technology innovation (Grant No.~JPMJSP2110).

\section*{Author declarations}
\subsection*{Conflict of Interest}
The authors have no conflicts to disclose.

\section*{Data availability}
The data that support the findings of this study are available from the corresponding author upon reasonable request.



\bibliography{tanimura_publist,AB,references}
\end{document}